\documentclass[twocolumn]{jpsj2tc} 
\title{Spectral filtering in quantum Y-junction}

\author{
Taksu \textsc{Cheon}$^1\!\!$
\thanks{Email address: taksu.cheon@kochi-tech.ac.jp}
Pavel \textsc{Exner}$^{2,3}\!\!$
\thanks{Email address: exner@ujf.cas.cz}
Ond{\v r}ej \textsc{Turek}$^{2,4}\!\!$
\thanks{Email address: turekond@fjfi.cvut.cz}
}
\inst{
$^1$Laboratory of Physics, Kochi University of Technology,
Tosa Yamada, Kochi 782-8502, Japan\\
$^2$Doppler Institute, Czech Technical University, 
Brehova 7, 11519 Prague , Czech Republic\\
$^3$Nuclear Physics Institute, Czech Academy of Sciences, 
25068 {\v R}e{\v z} u Prahy, Czech Republic\\
$^4$Department of Mathematics, Czech Technical University, 
Trojanova 13, 12000 Prague, Czech Republic 
}
\abst{
We examine scattering properties of singular vertex of
degree $n=2$ and $n=3$, taking advantage of a new form
of representing the vertex boundary condition, which has been
devised to approximate a singular vertex with finite potentials.   
We show that proper identification of $\delta$ and $\delta'$ components
in the connection condition between outgoing lines enables the designing
of quantum spectral branch-filters. 
}

\kword{quantum graph, singular vertex, quantum wire, spectral filtering}

\begin{document}
\maketitle

\section {Introduction}
The quantum graph is an abstract mathematical model of 
single-electron 
quantum device made up of interconnected one-dimensional lines,
in which quantum particles propagate \cite{EKST08}.
Fundamental element of quantum graph is the star graph, or the singular vertex
of degree $n$, which is a single node where 
$n$ outgoing half-lines are connected.
Although the general mathematical characterization of a singular vertex
in terms of parameter space of unitary group $U(n)$ has been
there for some time \cite{KS99, Ha00, KS00, FT00, TF01}, the analysis of its physical 
contents other than the simplest case of $n=2$ is still missing.
In this article, we address the problem of making sense of $U(n)$ parameter space
by examining the basic and simplest example of $n=3$ singular vertex,
or Y-junction, in detail.
We show that the recent work on the approximation of singular vertex by
finite potentials supplies the basis for our analysis.
Central to the physical understanding of singular vertex is the realization
that a connection between each pair of outgoing lines can be classified
by its $\delta$ and $\delta'$ contents supplemented 
by ``magnetic'' phase change \cite{CET09}.
We show that this classification leads directly to the spectral filtering property
between the pair of lines, enabling us to design the spectral branching filter 
using quantum Y-junction.
%

\section {Reduction of boundary matrices}
%
Consider a quantum particle on a star graph with a single node and $n$ 
half lines. The system is specified by boundary conditions
that have in general the following structure, 
\begin{eqnarray}
\label{e01}
A \Psi + B \Psi^\prime = 0,
\end{eqnarray}
where $A$ and $B$ are matrices $n\times n$ which must satisfy certain conditions, 
and $\Psi$, $\Psi'$ are the state vectors given by
\begin{eqnarray}
\label{e02}
\Psi  = \begin{pmatrix} \varphi_1 \cr \vdots\cr  \varphi_n \end{pmatrix} ,
\quad
\Psi^\prime  = \begin{pmatrix} \varphi'_1 \cr \vdots\cr  \varphi'_n \end{pmatrix}.
\end{eqnarray}
For simplicity of the notation, 
we have dropped the $x$ location when it is $x=0$,
i.e. we use 
$\varphi_i$, $\varphi'_i$ in place of $\varphi_i(0)$, $\varphi'_i(0)$. 
In this paper we start from the form of $A$, $B$ 
that we have devised in our previous work \cite{CET09} and 
where the crucial numbers are the ranks of the matrices $A$ and $B$
which we denote here $r_A=\mathrm{rank}(A)$ and $r_B=\mathrm{rank}(B)$. 
We can transform the $n\times n$ matrices $A$ 
and $B$ to the following {\it ST form};
\begin{eqnarray}
\label{e011}
A = - \begin{pmatrix} S & 0 \cr -T^\dagger & I \end{pmatrix},
\quad
B = \begin{pmatrix} I & T \cr 0 & 0 \end{pmatrix} ,
\end{eqnarray}
with $r_B \times r_B$ Hermitian matrix $S$ and $r_B \times (n-r_B)$
complex matrix $T$. 
The identity submatrix $I$
is  understood as having proper dimensions, 
namely $r_B \times r_B$ in $B$ and $(n-r_B) \times (n-r_B)$ in $A$.
If we denote the rank of $S$ as $r_S$, we obviously have
$0 \le r_S \le r_B$, and moreover,
\begin{eqnarray}
\label{e015}
r_A+r_B=n+r_S,
\end{eqnarray}
which comes in handy to us later on. 

Let us consider the scattering solution for incoming wave entering 
from $j$-th line with the wave number $k$;
\begin{eqnarray}
\label{e04}
\varphi_i^{(j)} (x_i)
&& \!\!\!\!\!\!\!\!\!\!\!\! 
= e^{-{\textrm i} k x_i} + {\cal R}_{i} e^{{\textrm i} k x_i} \ \ \ (i=j) , 
\nonumber \\
&& \!\!\!\!\!\!\!\!\!\!\!\! 
= {\cal T}_{i j}\, e^{{\textrm i} k x_i} \qquad\qquad\ \  (i \ne j),
\end{eqnarray}
where ${\cal R}_i$ represents the reflection amplitude for $i$-th line,
and ${\cal T}_{ij}$ the transmission amplitude from $j$-th to $i$-th line.
From the vectors $\Psi^{(j)}$ and $\Psi'^{(j)}$ made from
$\varphi^{(j)}_i$ and $\varphi'^{(j)}_i$ respectively,  we can construct matrices 
\begin{eqnarray}
\label{e044}
&&\!\!\!\!\!\!\!\!
(\Psi^{(1)} \cdots \Psi^{(n)}) = {\cal S}(k) + I,
\nonumber \\
&&\!\!\!\!\!\!\!\!
(\Psi'^{(1)} \cdots \Psi'^{(n)}) = {\textrm i}k ( {\cal S}(k) - I ) .
\end{eqnarray}
where the scattering matrix ${\cal S}(k)$ (which is not to be confused
with the sub-matrix $S$ appearing in (\ref{e011})) is given by
\begin{eqnarray}
\label{e05}
{\cal S}(k)=\begin{pmatrix} 
 {\cal R}_{1}(k)   & {\cal T}_{12}(k) & \cdots & {\cal T}_{1n}(k) \cr 
 {\cal T}_{21}(k) & {\cal R}_{2}(k)   & \cdots & {\cal T}_{2n}(k) \cr
 \vdots & & & \vdots \cr
 {\cal T}_{n1}(k) & {\cal T}_{n2}(k)& \cdots & {\cal R}_{n}(k) \end{pmatrix} .
\end{eqnarray}
From (\ref{e01}), we obtain
\begin{eqnarray}
\label{e06}
{\cal S}(k) = - \frac{1}{A + {\textrm i} k B} (A - {\textrm i} k B) ,
\end{eqnarray}
where $\frac{1}{M}$ represents the inverse matrix of $M$.

A vertex coupling can be also described by boundary conditions formulated as ${\bar A}\Psi+ {\bar B}\Psi'=0$
for
\begin{eqnarray}
\label{e07}
{\bar A}  =  \begin{pmatrix} I & \bar{T} \cr 0 & 0 \end{pmatrix} , 
\ \ 
{\bar B} =  -\begin{pmatrix} \bar{S} & 0 \cr -\bar{T}^\dagger & I \end{pmatrix};
\end{eqnarray}
this will be called a {\it reverse ST form}. 
It is obvious that for a given vertex coupling the matrices $A$ and $\bar{A}$ differ, as well as $B$, $\bar{B}$. And conversely, a simple interchange of $A$ and $B$ in \eqref{e02}, namely $B\Psi+ A\Psi'=0,$ leads to boundary conditions that correspond to a different system; this system may be considered as a counterpart of the original one. 
Let us examine how the scattering matrices are related in this case:
%
\begin{eqnarray}
\label{e08}
{\cal S}_d(k) 
&&\!\!\!\!\!\!\!\!\!\!\!\!
= - \frac{1}{ B + {\textrm i}k  A} ( B -{\textrm i}k  A)
\nonumber \\
&&\!\!\!\!\!\!\!\!\!\!\!\!
=  \frac{1}{A + \frac{{{\textrm i}}}{-k} B} 
(A - \frac{{\textrm i}}{-k} B) 
=- {\cal S}(-1/k).
\end{eqnarray}
This formula signifies a {\it high-low wave number duality} $k \leftrightarrow -1/k$
between the scattering matrix ${\cal S}(k)$ of system described by the
ST form and ${\cal S}_d(k)$ of its counterpart.

We now consider a single system and two its characterizations: one by the ST form
$A\Psi+ B\Psi'=0$ with (\ref{e011}), one by the reverse ST form ${\bar A}\Psi+ {\bar B}\Psi'=0$ with \eqref{e07}.
%
%
Although the matrices $A$ and $\bar{A}$ are very different, as well as $B$, $\bar{B}$, it naturally holds 
$\mathrm{rank}(A)=\mathrm{rank}({\bar A})$ and 
$\mathrm{rank}(B)=\mathrm{rank}({\bar B})$, which, because of (\ref{e015}), 
further leads to $\mathrm{rank}(S)=\mathrm{rank}({\bar S})$.
In other words, the quantity
$r_S=r_A+r_B-n$ is a characteristic
number of a system, that is independent of the representation.

\section {Scattering matrices and boundary conditions: n=2 case}
%
We start by examining the known case of $n=2$, namely, the 
{\it  point interaction on a line}, in order to see the effectiveness 
of our ST form in identifying the physical content of the singular vertex.

\subsection {\bf rank($B$)=0, rank($A$)=2}
For this case, the first condition $\mathrm{rank}(B)=0$
automatically guarantees the second condition $\mathrm{rank}(A)=2$.
We have the equation
\begin{eqnarray}
\label{e11}
\Psi=0,
\end{eqnarray}
which determines disjoint Dirichlet boundaries $\varphi_1$ $=\varphi_2$ $=0$.
\subsection {rank($B$)=1}
Suppose we now have $\mathrm{rank}(B)=1$.
The relation (\ref{e015}) reads $\mathrm{rank}(A)=\mathrm{rank}(S)+1$.
There are two possibilities.

\subsubsection {\bf rank($B$)=1, rank($A$)=1}
This corresponds to $\mathrm{rank}(S)=0$. We have the equation
\begin{eqnarray}
\label{e116}
\begin{pmatrix} 1 & t \cr 0 & 0 \end{pmatrix}
\begin{pmatrix} \varphi'_1 \cr \varphi'_2 \end{pmatrix}
= \begin{pmatrix} 0 & 0 \cr -t^* & 1 \end{pmatrix}
\begin{pmatrix} \varphi_1 \cr \varphi_2 \end{pmatrix},
\end{eqnarray}
which is the pure F{\"u}l{\"o}p-Tsutsui scale invariant
boundary condition \cite{FT00},
$t^*\varphi_1=\varphi_2$ and $\varphi'_1=-t \varphi'_2$.

\subsubsection {\bf rank($B$)=1, rank($A$)=2}
This corresponds to $\mathrm{rank}(S)=1$.
We have, in this case, the form
\begin{eqnarray}
\label{e12}
\begin{pmatrix} 1 & t \cr 0 & 0 \end{pmatrix}
\begin{pmatrix} \varphi'_1 \cr \varphi'_2 \end{pmatrix}
= \begin{pmatrix} s & 0 \cr -t^* & 1 \end{pmatrix}
\begin{pmatrix} \varphi_1 \cr \varphi_2 \end{pmatrix},
\end{eqnarray}
with a non-zero real number $s$ and a complex number $t$.
With $t=1$, we have $\varphi'_1 + \varphi'_2$ $= s \varphi_1 = s \varphi_2$,
which is nothing but the $\delta$ interaction with strength $s$.
(Note the outgoing directions for all $x_i$s.)

In general, the case $\mathrm{rank}(B)=1$ is understood as the combination
of $\delta$ and F{\" u}l{\" o}p-Tsutsui interactions.
This is evident from the transmission amplitude
\begin{eqnarray}
\label{e121}
{\cal T}_{12}(k) = \frac{2 k t }{k (1+t^*t)+{\textrm i}s} ,
\end{eqnarray}
whose characteristic length scale is $(1+t^*t)/s$.  Inverse of this length scale
divides the wave number into two regions.  We find 
the low wave number blockade ${\cal T}_{12}(0)=0$ and
high wave number transparency ${\cal T}_{12}(\infty)=\frac{2t}{1+t^*t}$ 
which becomes the perfect transparency ${\cal T}_{12}(\infty)=1$ for $t=1$.

\subsection {rank($B$)=2}
We have the form
\begin{eqnarray}
\label{e13}
\begin{pmatrix} \varphi'_1 \cr \varphi'_2 \end{pmatrix}
= \begin{pmatrix} s_{11} & s_{12} \cr s^*_{12} & s_{22} \end{pmatrix}
\begin{pmatrix} \varphi_1 \cr \varphi_2 \end{pmatrix}.
\end{eqnarray}
From the relation (\ref{e015}), we obtain
$\mathrm{rank}(A)=\mathrm{rank}(S)$, which leaves us with
three possibilities $\mathrm{rank}(A)=0$, $1$ and $2$.

\subsubsection {\bf rank($B$)=2, rank($A$)=0}
This corresponds to $\mathrm{rank}(S)=0$, and we have the equation
\begin{eqnarray}
\label{e136}
\Psi'=0,
\end{eqnarray}
representing disjoint Neumann boundaries $\varphi'_1$ $=\varphi'_2$ $=0$.

\subsubsection {\bf rank($B$)=2, rank($A$)=1}
When the rank of the matrix $A$ is one,
we can re-parametrize the above equation as
\begin{eqnarray}
\label{e14}
\begin{pmatrix} \varphi'_1 \cr \varphi'_2 \end{pmatrix}
= \begin{pmatrix} s & c s \cr c^* s & c^* c s \end{pmatrix}
\begin{pmatrix} \varphi_1 \cr \varphi_2 \end{pmatrix}
\end{eqnarray}
with a real number $s$ and a complex number $c$.
Multiplying the both sides by 
\begin{eqnarray}
\label{e15}
\begin{pmatrix} 1/s & 0 \cr -c^* & 1 \end{pmatrix},
\end{eqnarray}
we obtain the reverse ST form,
\begin{eqnarray}
\label{e16}
\begin{pmatrix} {\bar s} & 0 \cr -{\bar t}^* & 1 \end{pmatrix}
\begin{pmatrix} \varphi'_1 \cr \varphi'_2 \end{pmatrix}
= \begin{pmatrix} 1 & {\bar t} \cr 0 & 0 \end{pmatrix}
\begin{pmatrix} \varphi_1 \cr \varphi_2 \end{pmatrix},
\end{eqnarray}
with ${\bar s}=1/s$ and ${\bar t} = c$,
signifying the {\it pure $\delta'$ interaction}
amended by the F{\" u}l{\" o}p-Tsutsui scaling.
The transmission amplitude,
\begin{eqnarray}
\label{e17}
{\cal T}_{12}(k) = \frac{ -2 {\bar t}  }{ (1+{\bar t}^*{\bar t})-{\textrm i} k{\bar s}  } ,
\end{eqnarray}
shows both the high-wave number blockade, ${\cal T}_{12}(\infty)=0$, 
and low-wave number pass filtering behavior, 
${\cal T}_{12}(0)=\frac{-2{\bar t}}{1+t^*t}$.
Obviously, this is a dual partner of previous example of pure $\delta$
connection.

\subsubsection {\bf rank($B$)=2, rank($A$)=2}
When the rank of the matrix $A$ is two, we have the generic connection 
condition for a quantum particle residing on two joint lines, namely the
combinations of $\delta$ and $\delta'$ interactions.
This can be seen from the low-wave number and high-wave number blockade
behavior
\begin{eqnarray}
\label{e18}
{\cal T}_{12}(k) = \frac{2 k s_{12}  }{i k^2  - k \, {\rm tr}[S] -{\textrm i} \det[S]} .
\end{eqnarray}

In summary, for the case of $n=2$, 
the rank of the matrices $A$ and $B$, and resultantly, that of $S$, 
are the determining factors of physical contents of point interactions.  
\section {Scattering matrices and boundary conditions: n=3 case}
We now examine the quantum Y-junction, namely, the singular vertex of $n=3$.  
We shall show that the concept of ``$\delta$-like'' and ``$\delta'$-like'' couplings
can be established between each pair of lines outgoing from the singular vertex.

In idealized limit, two lines $i$ and $j$ are identified as having 
``pure $\delta$-like'' connections when we have 
\begin{eqnarray}
\label{e301}
{\cal T}_{ij}(0) = 0 , \quad {\rm and} 
\quad 
{\cal T}_{ij}(k) = Const. \ (k\to\infty).
\end{eqnarray}
Conversely, $i$ and $j$ are identified as `` pure $\delta'$-like''  if we have
\begin{eqnarray}
\label{e302}
{\cal T}_{ij}(0) = Const. \ (k \to 0), \quad {\rm and} \quad 
{\cal T}_{ij}(\infty) = 0.
\end{eqnarray}
Since the quantum flux can circumvent  direct blocking between $i$ and $j$
through indirect path $i \to k \to j$, strict conditions ${\cal T}_{ij}(0)=0$ 
for $\delta$-like and ${\cal T}_{ij}(\infty)=0$ for $\delta'$-like connection 
are to be breached when other types of connections are present among 
other lines, and therefore, zeros for ${\cal T}_{ij}$ need to be
replaced by {\it small} number, ${\cal T}_{ij}\approx 0$ in above conditions. 
General characterization of pure $\delta$-like connection as high-pass 
frequency filter, and pure $\delta'$-like connection low-pass filter
is still valid.  

%
As in the case of $n=2$, we classify the boundary condition
according to the ranks of matrices $A$ and $B$.

\subsection {\bf rank($B$)=0, rank($A$)=3}
The first condition automatically ensures the second.
We again have disjoint condition
\begin{eqnarray}
\label{e31}
\Psi=0 ,
\end{eqnarray}
which is disconnected Dirichlet boundaries $\varphi_1$ $=\varphi_2$
$=\varphi_3$ $=0$.

\subsection {rank($B$)=1}

With this condition,
the relation (\ref{e015}) now reads $\mathrm{rank}(A)=\mathrm{rank}(S)+2$.
There are two possibilities, $\mathrm{rank}(A)=2$ and $3$.

\subsubsection {\bf rank($B$)=1, rank($A$)=2}
This corresponds to $\mathrm{rank}(S)=0$, and we have the equation
\begin{eqnarray}
\label{e316}
\begin{pmatrix} 1 & t_2& t_3 \cr 0 & 0 & 0 \cr 0 & 0 & 0 \end{pmatrix}
\begin{pmatrix} \varphi'_1 \cr \varphi'_2 \cr \varphi'_3 \end{pmatrix}
= \begin{pmatrix} 0 & 0 & 0 \cr -t_2^* & 1 & 0 \cr   -t_3^* & 0 & 1\end{pmatrix}
\begin{pmatrix} \varphi_1 \cr \varphi_2 \cr  \varphi_3 \end{pmatrix},
\end{eqnarray}
which is $n=3$ version of pure scale invariant F{\"u}l{\"o}p-Tsutsui boundary 
condition, given by
$t_2^* t_3^*\varphi_1=t_3^*\varphi_2=t_2^*\varphi_3$ and 
$\varphi'_1+t_2 \varphi'_2+t_3 \varphi'_3=0$.

\subsubsection {\bf rank($B$)=1, rank($A$)=3}
This case corresponds to $\mathrm{rank}(S)=1$.
We have
\begin{eqnarray}
\label{e32}
\begin{pmatrix} 1 & t_2 & t_3 \cr 0 & 0 & 0 \cr 0 & 0 & 0 \end{pmatrix}
\begin{pmatrix} \varphi'_1 \cr \varphi'_2 \cr \varphi'_3 \end{pmatrix}
= \begin{pmatrix} s & 0  & 0\cr -t_2^* & 1 & 0 \cr -t_3^* & 0 & 1 \end{pmatrix}
\begin{pmatrix} \varphi_1 \cr \varphi_2 \cr \varphi_3 \end{pmatrix},
\end{eqnarray}
with non-zero real number $s$.
With $t_2= t_3=1$, we have $\varphi'_1 + \varphi'_2 + \varphi'_3$ 
$= s \varphi_1 = s \varphi_2 = s \varphi_3$,
which is the $n=3$ generalization of 
{\it pure $\delta$ potential connection conditions} \cite{Ex96} 
between all half lines.
With general $t_2$ and $t_3$,
F{\" u}l{\" o}p-Tsutsui scalings $t^*_2$, $t^*_3$ and $t^*_2/t^*_3$ are introduced
on $\varphi_2/\varphi_1$, $\varphi_3/\varphi_1$ and on $\varphi_2/\varphi_3$,
respectively.
%
\begin{figure}[h]
  \centering
  \includegraphics[width=3cm]{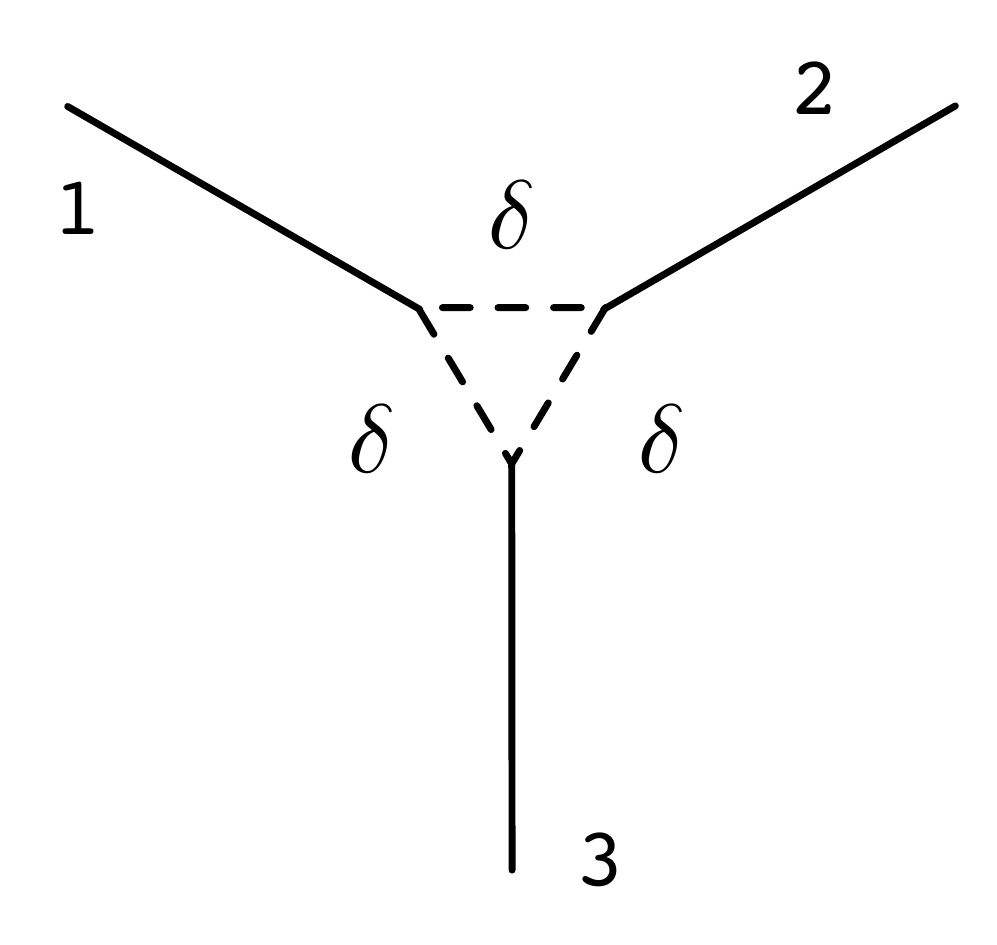}
  \caption{Pure $\delta$ type connection between all lines, obtained from ST form
  with ${\rm rank}(B)=1$ and ${\rm rank}(A)=3$.}
  \label{fig:ddd}
\end{figure}
\begin{figure}[h]
  \centering
  \includegraphics[width=9cm]{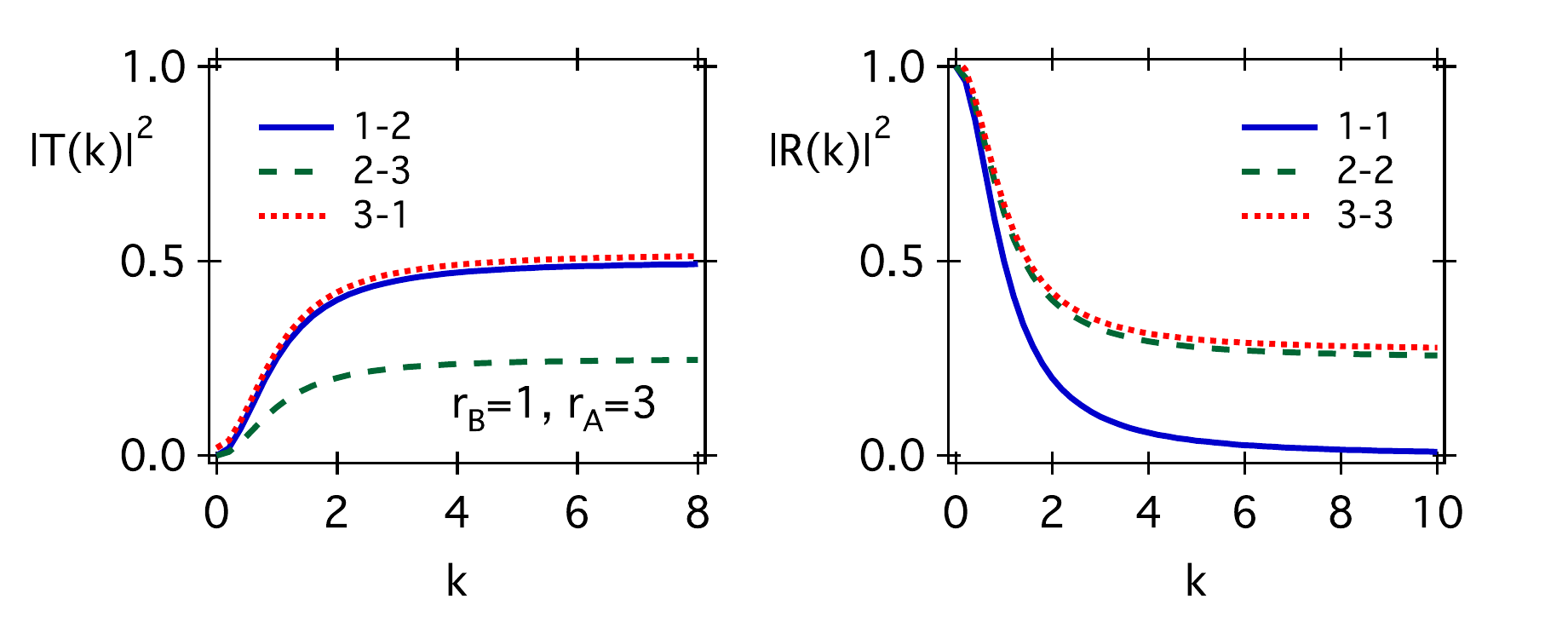}
  \caption{Transmission and reflection probabilities
   for Y-junction with pure $\delta$ type connection between all lines.
   In the left side, solid line represents $|{\cal T}_{12}(k)|^2$, 
   dashed line $|{\cal T}_{23}(k)|^2$, and dotted line $|{\cal T}_{31}(k)|^2$.
   In the right, solid line represents $|{\cal R}_{1}(k)|^2$, 
   dashed line $|{\cal R}_{2}(k)|^2$, and dotted lin $|{\cal R}_{3}(k)|^2$. 
   Parameter values $t_2=t_3=1/\sqrt{2}$, $s=2$ are used in (27).
   Two identical lines are drawn with slight offsets for better viewing.}
  \label{fig:m1}
\end{figure}
The transmission amplitudes for this case are given by
\begin{eqnarray}
\label{e321}
&&
{\cal T}_{31}(k) = \frac{2 t_3^*k  }{{\textrm i} s + (1+t_2^*t_2+t_3^*t_3) k}  ,
\nonumber \\
&&
{\cal T}_{12}(k) = \frac{2 t_2 k  }{{\textrm i} s + (1+t_2^*t_2+t_3^*t_3) k}  ,
\nonumber \\
&&
{\cal T}_{23}(k) = \frac{2 t_2^* t_3 k  }{{\textrm i} s + (1+t_2^*t_2+t_3^*t_3) k} .
\end{eqnarray}
which has  the length scale $(1+t_2^*t_2+t_3^*t_3)/s$.
Below this length scale, the transmission coefficients 
show the high-wave number pass filtering behavior 
\begin{eqnarray}
\label{e322}
&&
{\cal T}_{ij}(0) = 0 , \quad  {\cal T}_{ij}(k) = Const. \ \ {\rm as} \ k \to 0,
\end{eqnarray}
which is a hallmark of pure $\delta$ connections  between all branches
(See Figs. 1 and 2).

\subsection {rank($B$)=2}
The ST form $B\Psi' = - A\Psi$ now reads
\begin{eqnarray}
\label{e33}
\begin{pmatrix} 1 & 0 & t_1 \cr 0 & 1 & t_2 \cr 0 & 0 & 0 \end{pmatrix}
\begin{pmatrix} \varphi'_1 \cr \varphi'_2 \cr \varphi'_3 \end{pmatrix}
= \begin{pmatrix} s_{11} & s_{12} & 0\cr s^*_{12} & s_{22} & 0 \cr
 -t_1^* & -t_2^* & 1 \end{pmatrix}
\begin{pmatrix} \varphi_1 \cr \varphi_2 \cr \varphi_3 \end{pmatrix}.
\end{eqnarray}
The relation (\ref{e015}) becomes
$\mathrm{rank}(A)=1+\mathrm{rank}(S)$.
We have three possibilities:

\subsubsection {\bf rank($B$)=2, rank($A$)=1}
This corresponds to $\mathrm{rank}(S)=0$.  
We have $s_{11}=s_{12}=s_{22}=0$ in (\ref{e33}).  This situation represents a
scale invariant interaction between lines 1--3, described by $\varphi'_1=-t_3 \varphi'_3$, and a scale invariant interaction between lines 2--3, described by
$\varphi'_2=-t_3 \varphi'_3$.

\subsubsection {\bf rank($B$)=2, rank($A$)=2}
Suppose that the rank of the sub-matrix $S$ is one, namely top two raws of 
the RHS are linearly dependent to each other.
We can write (\ref{e33}) in the form
\begin{eqnarray}
\label{e330}
\begin{pmatrix} 1 & 0 & t_1 \cr 0 & 1 & t_2 \cr 0 & 0 & 0 \end{pmatrix}
\begin{pmatrix} \varphi'_1 \cr \varphi'_2 \cr \varphi'_3 \end{pmatrix}
= \begin{pmatrix} s & c s & 0 \cr c^* s & c^* c s & 0\cr  -t_1^* & -t_2^* & 1 \end{pmatrix}
\begin{pmatrix} \varphi_1 \cr \varphi_2 \cr \varphi_3 \end{pmatrix}.
\end{eqnarray}
Interestingly, we can reverse the role of $A$ and $B$ in the following manner.
We now write (\ref{e330}) in the form
\begin{eqnarray}
\label{e331}
\begin{pmatrix} 1 & t_1 & 0 \cr 0 & 0 & 0 \cr 0 &  t_2 & 1 \end{pmatrix}
\begin{pmatrix} \varphi'_1 \cr \varphi'_3 \cr \varphi'_2 \end{pmatrix}
= \begin{pmatrix} s & 0 & c s \cr -t_1^* &1 & -t_2^* \cr c^* s & 0 & c^* c s \end{pmatrix}
\begin{pmatrix} \varphi_1 \cr \varphi_3 \cr \varphi_2 \end{pmatrix}.
\end{eqnarray}
Multiplying the both sides by 
\begin{eqnarray}
\label{e332}
\begin{pmatrix} 1/s & 0 & 0 \cr t_1^*/s & 1 & 0\cr -c^* & 0 & 1 \end{pmatrix},
\end{eqnarray}
we obtain a reverse ST form $-B \Psi' = A \Psi$ as
\begin{eqnarray}
\label{e333}
\begin{pmatrix} {\bar s} & {\bar c} {\bar s} & 0\cr {\bar c}^* {\bar s} 
& {\bar c}^* {\bar c} s & 0 \cr
 -{\bar t}_1^* & -{\bar t}_3^* & 1 \end{pmatrix}
\begin{pmatrix} \varphi'_1 \cr \varphi'_3 \cr \varphi'_2 \end{pmatrix}
= \begin{pmatrix} 1 & 0 & {\bar t}_1 \cr 0 & 1 & {\bar t}_3 \cr 0 & 0 & 0 \end{pmatrix} 
\begin{pmatrix} \varphi_1 \cr \varphi_3 \cr \varphi_2 \end{pmatrix},
\end{eqnarray}
with ${\bar s} = 1/s$,  ${\bar  c} = t_1$, ${\bar  t_1} = c$, 
and ${\bar t}_3=ct_1^*-t_2^*$.
Note that two forms (\ref{e330}) and (\ref{e333}) are dual to each other,
and that this case can be also viewed as 
having ${\rm rank}(A)=2$ and ${\rm rank}({\bar S})=1$, as well as
${\rm rank}(B)=2$ and ${\rm rank}(S)=1$.

It is instructive to look at the transmission amplitudes,
which, for this case, are given by
\begin{eqnarray}
\label{e334}
&&\!\!\!\!\!\!\!\!
{\cal T}_{31}(k) = \frac{2 t_1^* k + 2 {\textrm i} c^* s(c t_1^*-t_2^*)  }
{D_1 k +{\textrm i} s D_0}  ,
\nonumber \\
&&\!\!\!\!\!\!\!\!
{\cal T}_{12}(k) = \frac{-2 t_2^*t_1 k - 2 {\textrm i} c s   }
{D_1 k +{\textrm i} s D_0}  ,
\nonumber \\
&&\!\!\!\!\!\!\!\!
{\cal T}_{23}(k) = \frac{2 t_2 k - 2 {\textrm i} s (c^* t_1-t_2)  }
{D_1 k +{\textrm i} s D_0}  .
\end{eqnarray}
where we set
\begin{eqnarray}
\label{e3341}
&&\!\!\!\!\!\!\!\!
D_0 = 1+c^*c + (c t_1^*-t_2^*)(c^* t_1-t_2),
\nonumber \\
&&\!\!\!\!\!\!\!\!
D_1 = 1+t_1^*t_1+t_2^*t_2 .
\end{eqnarray}
Two special cases are noteworthy, at which we shall look in detail.
\begin{figure}[h]
  \centering
  \includegraphics[width=3cm]{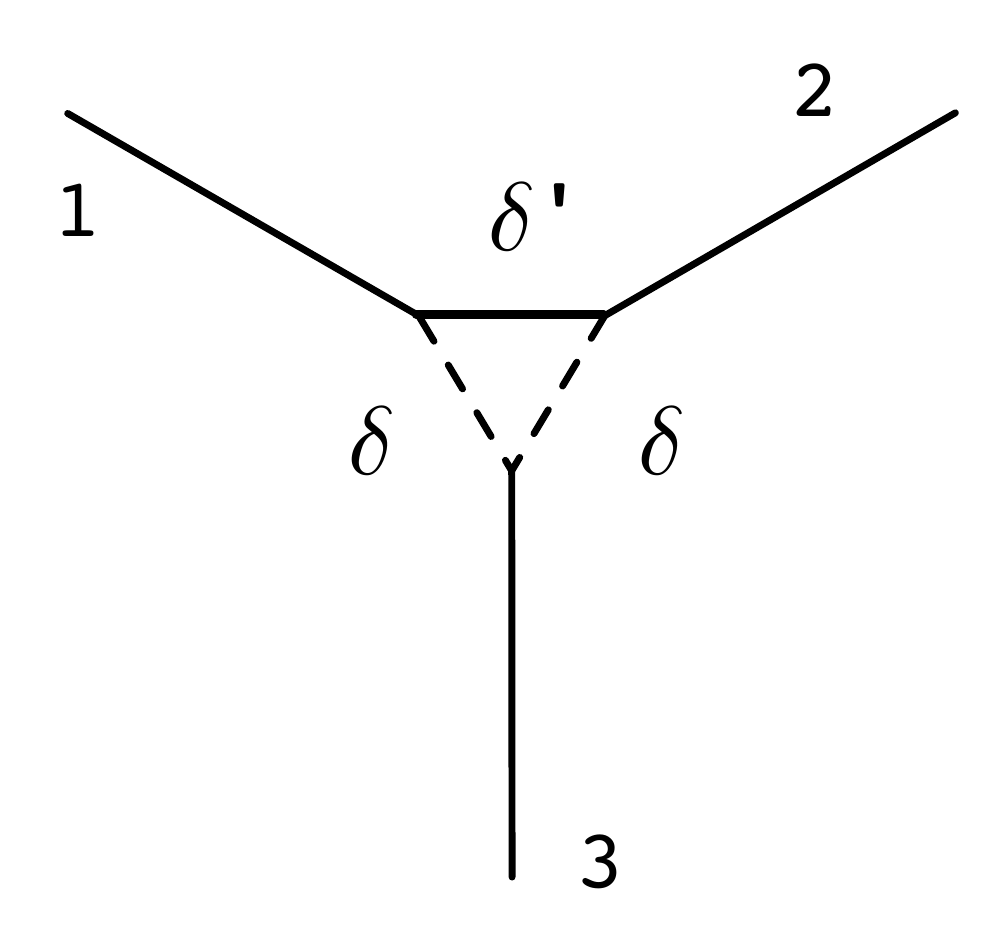}\quad
  \includegraphics[width=3cm]{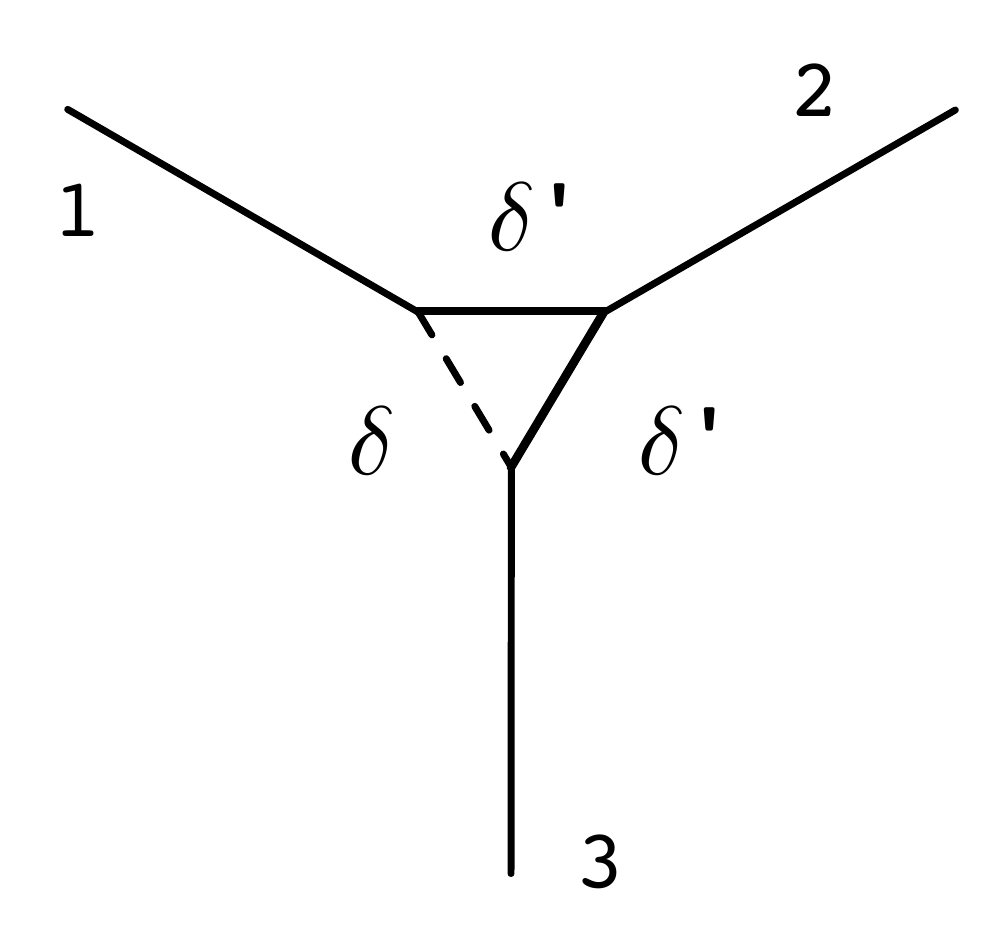}
  \caption{Mixed type vertex coupling obtained from ST form
  with  ${\rm rank}(B)=2$ and  ${\rm rank}(A)=2$. The
  $\delta$--$\delta$--$\delta'$ (left) and $\delta'$--$\delta'$--$\delta$ (right)
  type connection are obtained from conditions ${\bar t}_3=0$ 
  and $t_2=0$, respectively.}
  \label{fig:dp}
\end{figure}

\bigskip
\noindent{\it  4.3.2.1\ \ \ }{\bf $\delta$-$\delta$-$\delta'$ type}
\\ \indent
Let us suppose for now, that we have $c t_1^*-t_2^*(={\bar t}_3^*)=0$.
This results in ${\cal T}_{31}(0)={\cal T}_{23}(0)=0$, indicating the
presence of two pure $\delta$-like connections between lines $3-1$,
and between $2-3$.  When further condition $s \gg t_1^*t_1$ 
and  $c \ne 0$ are met,
we have $T_{12}(k)= Const.$ as $k\to 0$ and $T_{12}(\infty)\approx0$,
signifying the pure $\delta'$-like connection between lines $1-2$.
The same conclusion is drawn from
the consideration of connection conditions which reads
\begin{eqnarray}
\label{e33bc1}
&&\!\!\!\!\!\!\!\!\!\!\!\!\!\!\!\!\!\!\!\!\!\!\!\!
\frac{1}{t_1}\varphi'_1 + \varphi'_3 
=\frac{1}{t_2}\varphi'_2 + \varphi'_3 
=\frac{s}{t_1^* t_1} \varphi_3
\nonumber \\
&&\!\!\!\!
\varphi_3 = t_1^* \varphi_1+ t_2^* \varphi_2 
\nonumber \\
&&\!\!\!\!
\frac{1}{t_1}\varphi'_1 = \frac{1}{t_2} \varphi'_2.
\end{eqnarray}
The last equation, which is not independent of the first three, 
is shown to display the pure $\delta'$-like interaction
between the half lines $1$ and $2$,
ammended by the F{\" u}l{\" o}p-Tsutsui scaling by factor $t_1/t_2$. 
The first two equations clearly show the fact that the connections
between the half lines $2$ and $3$, and between $3$ and $1$ are
pure $\delta$-like (See Fig. 3, left, and Fig. 4).
\bigskip
\noindent{\it  4.3.2.2\ \ \ }{\bf $\delta'$-$\delta'$-$\delta$ type}
\\ \indent
Let us now suppose, in place of previous conditions, that we have
$t_2=0$ and $t_1\ne 0$.  We then have ${\cal T}_{12}(\infty)={\cal T}_{23}(\infty)=0$,
${\cal T}_{12}(k)=Const. \ne 0$ and ${\cal T}_{23}(k)=Const.\ne 0$ as we let $k\to 0$,
indicating the presence of two pure $\delta'$-like connections 
between lines $1-2$ and between $2-3$.
With further assumption $c  \ll1$, we have ${\cal T}_{31}(0)\approx 0$,  
signifying the pure $\delta$-like connection between lines $3-1$
(See Fig. 3, right, and Fig. 5).
These facts are again clearly
visible in the following expressions for the boundary condition;
\begin{eqnarray}
\label{e33bc2}
&&\!\!\!\!\!\!\!\!\!\!\!\!\!\!\!\!\!\!\!\!\!\!\!\!
\frac{1}{c}\varphi_1 + \varphi_2 
=\frac{1}{c t_1^*}\varphi_3 + \varphi_2 
=\frac{1}{s c^* c} \varphi'_2 ,
\nonumber \\
&&\!\!\!\!
\varphi'_2 = c^* \varphi'_1+ c^* t_1 \varphi'_3 ,
\nonumber \\
&&\!\!\!\!
t_1^* \varphi_1 = \varphi_3.
\end{eqnarray}

Thus we have shown that this case corresponds to a mixture of $\delta$ and
$\delta'$ connections including two pure connections
$\delta-\delta-\delta'$ and $\delta'-\delta'-\delta$ as two limiting cases.
\begin{figure}[h]
  \centering
  \includegraphics[width=9cm]{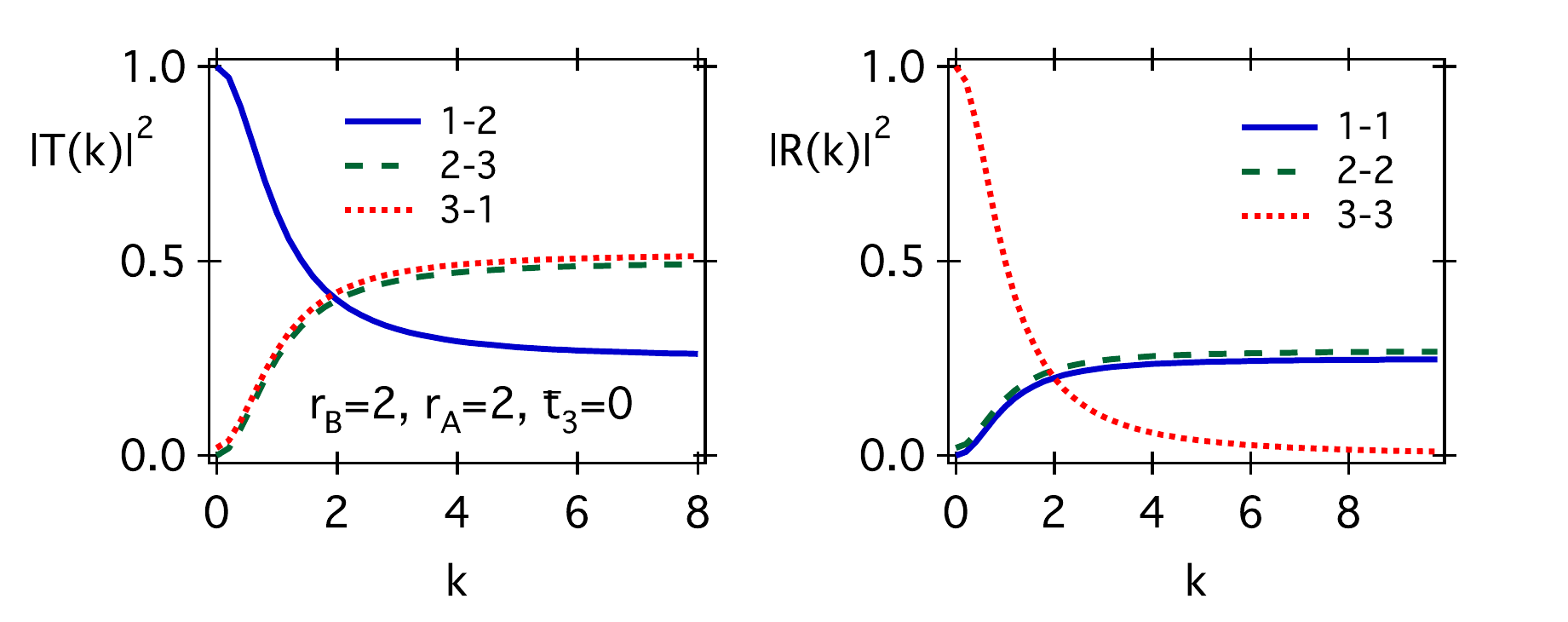}
  \caption{Transmission and reflection probabilities
   for Y-junction with $\delta$--$\delta$--$\delta'$  type connection.
   In the left side, solid line represents $|{\cal T}_{12}(k)|^2$, 
   dashed line $|{\cal T}_{23}(k)|^2$, and dotted line $|{\cal T}_{31}(k)|^2$.
   In the right, solid line represents $|{\cal R}_{1}(k)|^2$, 
   dashed line $|{\cal R}_{2}(k)|^2$, and dotted lin $|{\cal R}_{3}(k)|^2$.    
   Parameter values $t_1=t_2=1/\sqrt{2}$, $s_{11}=s_{12}=s_{22}=1$ 
   are used in \eqref{e33}.
   Two identical lines are drawn with slight offsets for better viewing.}
  \label{fig:m2s1}
\end{figure}
\begin{figure}[h]
  \centering
  \includegraphics[width=9cm]{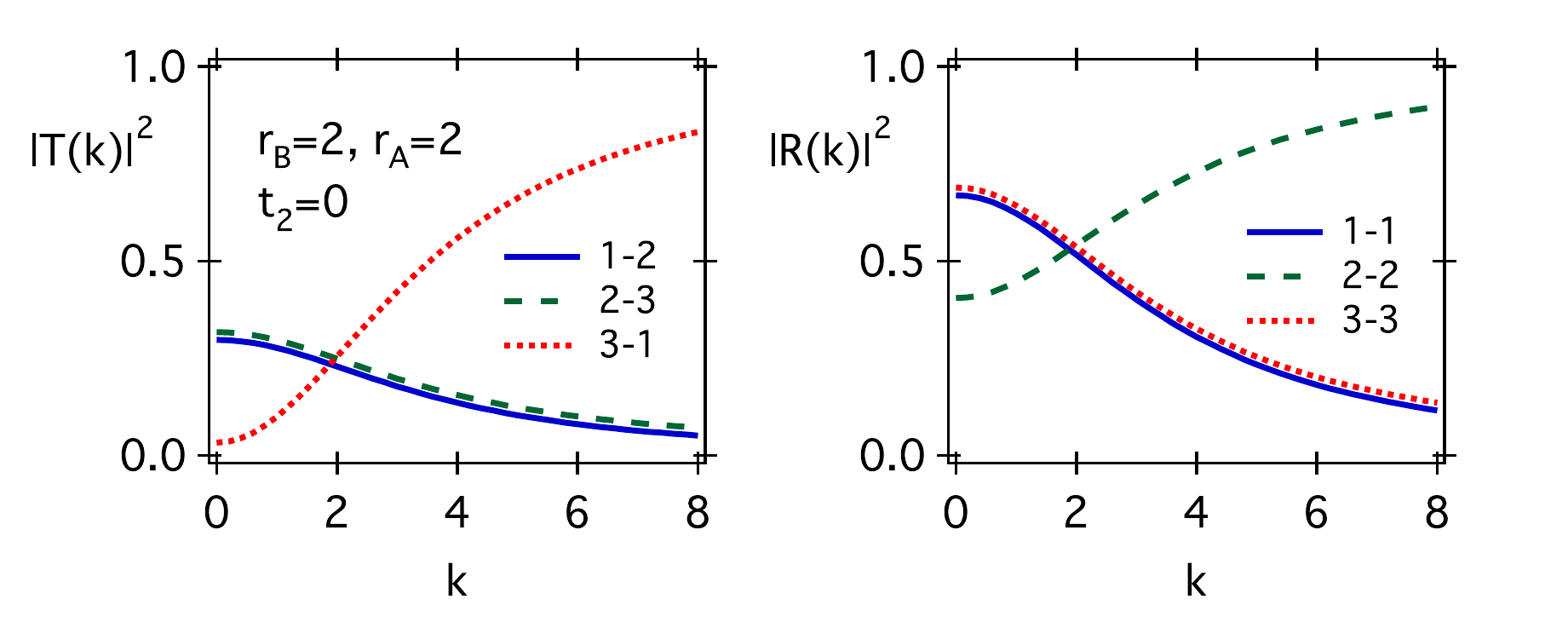}
  \caption{Transmission and reflection probabilities
   for Y-junction with $\delta$--$\delta'$--$\delta'$  type connection.
   Parameter values $t_1=1/3, t_2=0$, $s_{11}=6, s_{12}=2, s_{22}=2/3$
   are used in \eqref{e33}.}
  \label{fig:m2s1p}
\end{figure}

\subsubsection {\bf rank($B$)=2, rank($A$)=3}
When the rank of the matrix $A$ is three
(thus giving $\mathrm{rank}(S)=2$),
we have rather general combination
of $\delta$ and $\delta'$ interactions between each pair of half lines.
Let us look at the transmission amplitudes, which are given by
\begin{eqnarray}
\label{e335}
&&\!\!\!\!\!\!\!\!\!\!\!\!\!\!
{\cal T}_{31}(k) = \frac{2 t_1^* k^2 + 2 {\textrm i} (s_{22} t_1^*-s_{12}^* t_2^*) k  }
{k^2 E_2 +{\textrm i} k E_1 +E_0}   ,
\nonumber \\
&&\!\!\!\!\!\!\!\!\!\!\!\!\!\!
{\cal T}_{12}(k) = \frac{-2 t_2^* t_1 k^2 - 2 {\textrm i} s_{12} k  }
{k^2 E_2 +{\textrm i} k E_1 +E_0}   ,
\nonumber \\
&&\!\!\!\!\!\!\!\!\!\!\!\!\!\!
{\cal T}_{23}(k) = \frac{2 t_2 k^2 - 2 {\textrm i} (s_{12}^* t_1-s_{11}t_2) k  }
{k^2 E_2 +{\textrm i} k E_1 +E_0}  ,
\end{eqnarray}
where we set
\begin{eqnarray}
\label{e3351}
&&\!\!\!\!\!\!\!\!\!\!\!\!\!\!
E_0 = -\det[S]  ,
\nonumber \\
&&\!\!\!\!\!\!\!\!\!\!\!\!\!\!
E_1 = {\rm tr}[S] +  
s_{22} t_1^* t_1 - s_{12} t_1^* t_2 - s_{12}^* t_2^* t_1+ s_{11} t_2^* t_2
\nonumber \\
&&\!\!\!\!\!\!\!\!\!\!\!\!\!\!
E_2 = 1+t_1^*t_1+t_2^*t_2 .
\end{eqnarray}
The guaranteed presence of $\delta$-like connection between all lines can be 
seen from the zero energy blockade ${\cal T}_{ij}(0)=0$ for all $i$ and $j$.
The presence or absence of $\delta'$-like component is controlled by
$t_i$ since we have ${\cal T}_{31}(\infty)\propto t_1^* $, 
${\cal T}_{12}(\infty)\propto t_2^*t_1$
and ${\cal T}_{23}(\infty)\propto t_2 $.
A numerical example of this case is shown in Fig. 6.
\begin{figure}[h]
  \centering
  \includegraphics[width=9cm]{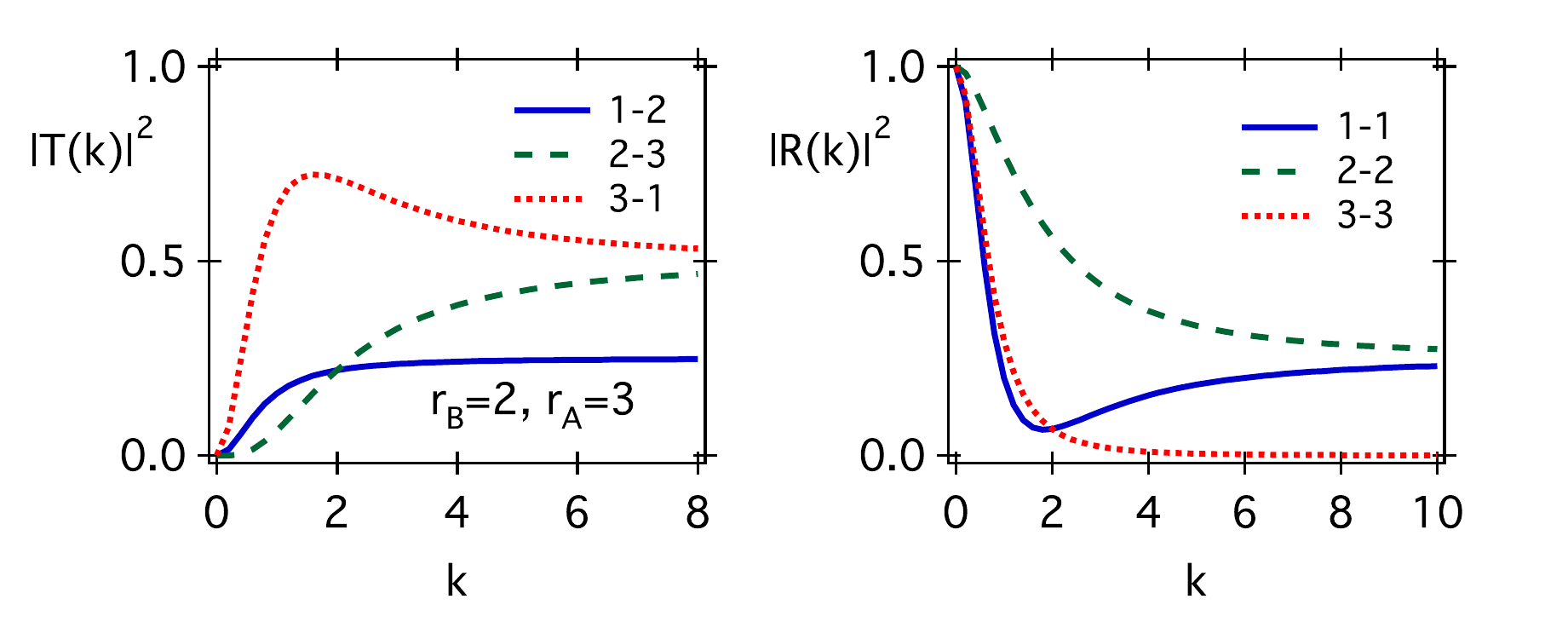}
  \caption{Transmission and reflection probabilities for Y-junction 
  with ${\rm rank}(B)=2$, ${\rm rank}(A)=3$.
  Parameter values $t_1=t_2=1/\sqrt{2}$, $s_{11}=s_{12}=1$, $s_{22}=-2$
  are used in (29).}
  \label{fig:m2s2}
\end{figure}

\subsection {rank($B$)=3}
We have the ST form
\begin{eqnarray}
\label{e34}
\begin{pmatrix} \varphi'_1 \cr \varphi'_2 \cr \varphi'_3 \end{pmatrix}
= \begin{pmatrix} s_{11} & s_{12} & s_{13} \cr
s^*_{12} & s_{22} & s_{23} \cr
s^*_{13} & s^*_{23} & s_{33} \end{pmatrix}
\begin{pmatrix} \varphi_1 \cr \varphi_2 \cr \varphi_3 \end{pmatrix}.
\end{eqnarray}
From (\ref{e011}), we have $A=S$, and thus
$\mathrm{rank}(A)=\mathrm{rank}(S)$.
We have four possibilities:

\subsubsection {\bf rank($B$)=3, rank($A$)=0}
This corresponds to $\mathrm{rank}(S)=0$, and the boundary condition
becomes
\begin{eqnarray}
\label{e347}
\Psi'=0
\end{eqnarray}
which is the disjoint Neumann condition $\varphi'_1$ $=\varphi'_2$
$=\varphi'_3$ $=0$.
\subsubsection {\bf rank($B$)=3, rank($A$)=1}
When the rank of the matrix $A$ is one, namely three rows of the RHS
are linearly dependent on each other, 
we have  
\begin{eqnarray}
\label{e341}
\begin{pmatrix} \varphi'_1 \cr \varphi'_2 \cr \varphi'_3 \end{pmatrix}
= \begin{pmatrix} s & c s & d s \cr
c^* s  & c^* c s & c^* d s\cr
d^* s & d^* c s & d^* d s \end{pmatrix}
\begin{pmatrix} \varphi_1 \cr \varphi_2 \cr \varphi_3 \end{pmatrix}.
\end{eqnarray}
Multiplying the both sides by 
\begin{eqnarray}
\label{e342}
\begin{pmatrix} 1/s & 0 & 0 \cr -c^* & 1 & 0\cr -d^* & 0 & 1 \end{pmatrix},
\end{eqnarray}
we arrive at a reverse ST form  as
\begin{eqnarray}
\label{e343}
\begin{pmatrix} {\bar s} & 0 & 0 \cr -c^* & 1 & 0\cr -d^* & 0 & 1 \end{pmatrix}
\begin{pmatrix} \varphi'_1 \cr \varphi'_2 \cr \varphi'_3 \end{pmatrix}
= \begin{pmatrix} 1 & c & d \cr 0 & 0 & 0 \cr 0 & 0 & 0 \end{pmatrix} 
\begin{pmatrix} \varphi_1 \cr \varphi_2 \cr \varphi_3 \end{pmatrix},
\end{eqnarray}
with ${\bar s} = 1/s$.
%
We have
$c^*  d^*  \varphi'_1 = d^* \varphi'_2 = c^*\varphi'_3$, and 
$\varphi_1 + c \varphi_2 + d \varphi_3 =  {\bar s} \varphi'_1$,
signifying the {\it generalized pure $\delta'$ interaction} \cite{Ex96}
ammended by the F{\" u}l{\" o}p-Tsutsui scaling.
This is also evident from the transmission amplitudes, which are given by
\begin{eqnarray}
\label{e345}
&&
{\cal T}_{31}(k) = \frac{ 2 d^* s  }
{-{\textrm i}k +s(1+c^*c +d^*d)}  ,
\nonumber \\
&&
{\cal T}_{12}(k) = \frac{ 2 c s  }
{-{\textrm i}k +s(1+c^*c +d^*d)}  ,
\nonumber \\
&&
{\cal T}_{23}(k) = \frac{ 2 c^* d s  }
{-{\textrm i}k +s(1+c^*c +d^*d)}  .
\end{eqnarray}
The formulae imply ${\cal T}_{ij}(\infty) = 0$ and ${\cal T}_{ij}(k) = Const.$ as $k \to 0$
(See Figs. 7 and 8).

\begin{figure}[h]
  \centering
  \includegraphics[width=3cm]{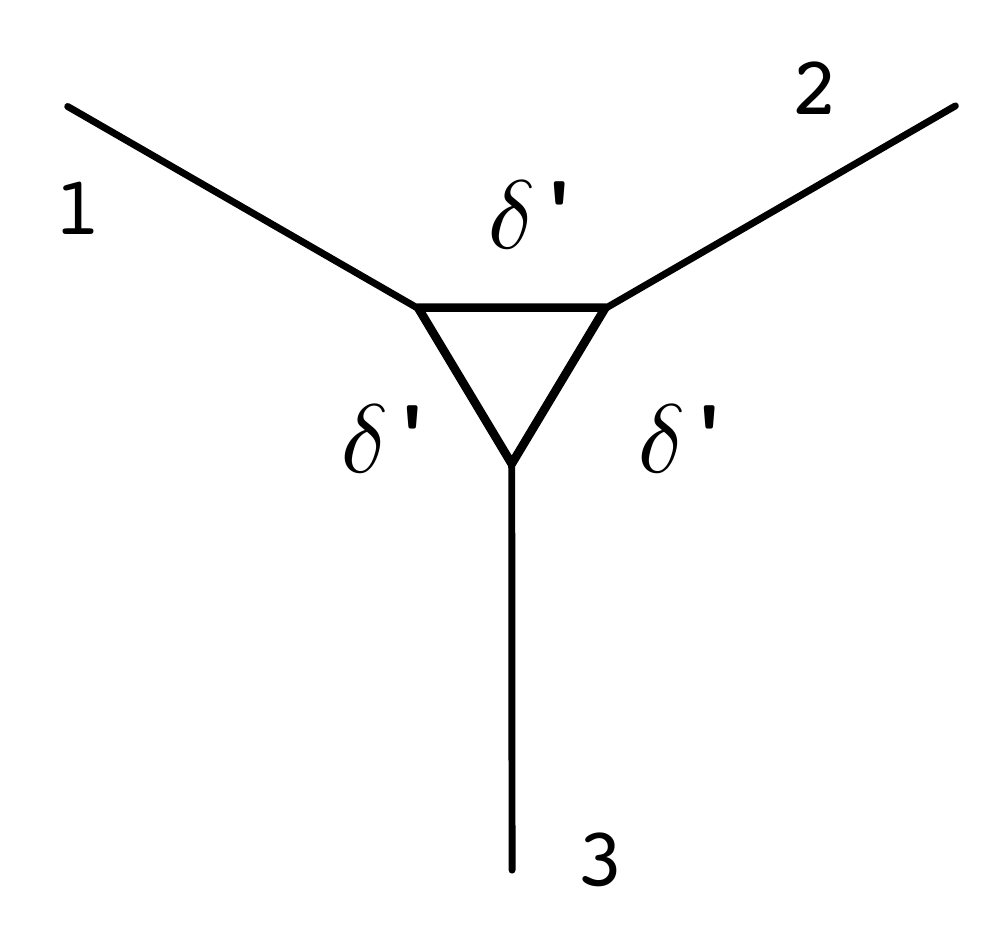}
  \caption{Pure $\delta'$ type connection between all lines, 
  obtained from ST form
  with ${\rm rank}(B)=3$ and ${\rm rank}(A)=1$.}
  \label{fig:ppp}
\end{figure}

\begin{figure}[h]
  \centering
  \includegraphics[width=9cm]{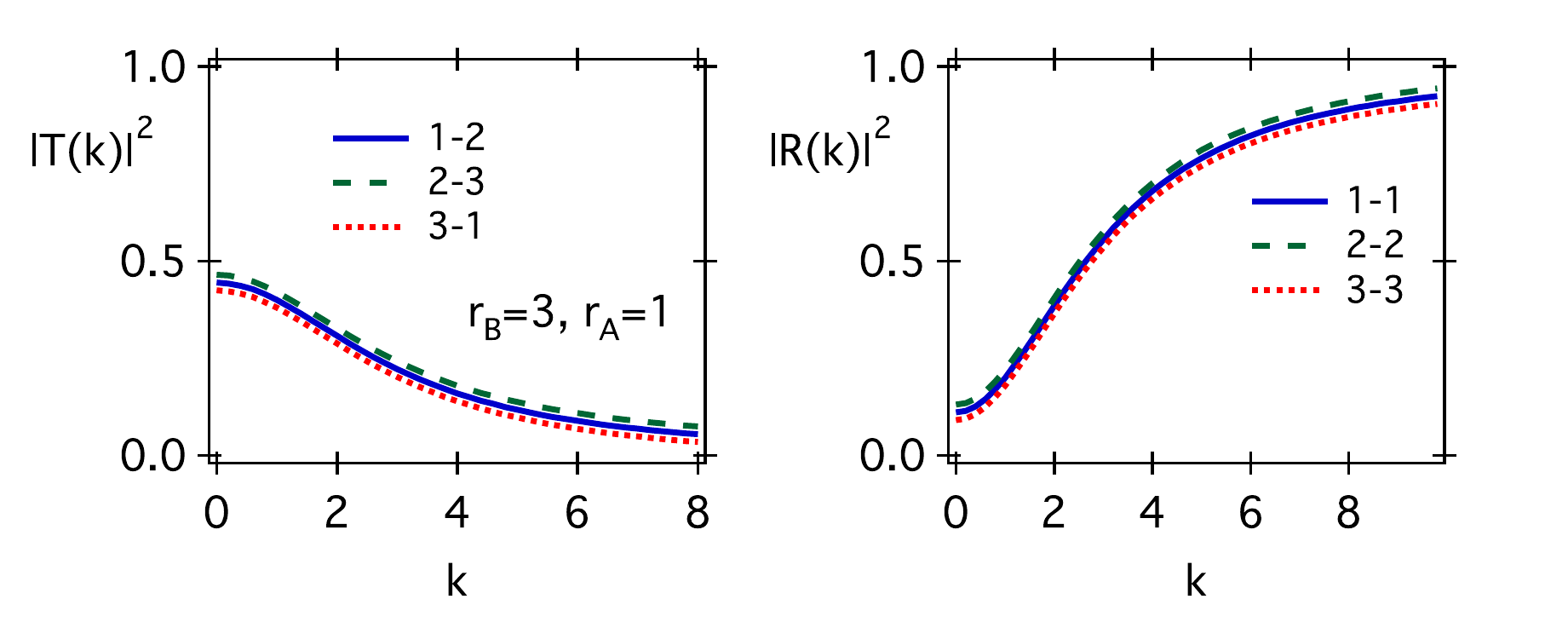}
  \caption{Transmission and reflection probabilities
   for Y-junction with pure $\delta'$ type connection between all lines.
   Parameter values $s_{11}=s_{12}=s_{13}=s_{22}=s_{23}=s_{33}=1$ 
   are used in \eqref{e34}.}
  \label{fig:m3s1}
\end{figure}

\subsubsection {\bf rank($B$)=3, rank($A$)=2}
When the rank of the matrix $A$ is two, and thus that of $S$ is two,
the last row of RHS of (\ref{e34})
is equal to some combination of the first two. 
We then have  
\begin{eqnarray}
\label{e351}
\begin{pmatrix} \varphi'_1 \cr \varphi'_2 \cr \varphi'_3 \end{pmatrix}
= \begin{pmatrix} s & q & \!\!\!\! c s+d q \cr
q^* & r  &  \!\!\!\! c q^*+d r \cr
c^* s + d^* q^* \! & \!  c^* q + d^*r \!\! &  f \end{pmatrix}
\begin{pmatrix} \varphi_1 \cr \varphi_2 \cr \varphi_3 \end{pmatrix}
\end{eqnarray}
with $f= c^*cs+ c^*dq + d^*cq^* + d^*dr$.
Multiplying both sides by 
\begin{eqnarray}
\label{e352}
\begin{pmatrix} r/(sr-q^*q) & -q/(sr-q^*q) & 0 \cr 
-q^*/(sr-q^*q) & s/(sr-q^*q) & 0\cr -c^* & -d^* & 1 \end{pmatrix},
\end{eqnarray}
we obtain a reverse ST form
\begin{eqnarray}
\label{e353}
\begin{pmatrix} {\bar  s} & {\bar q} & 0 \cr {\bar q}^* & {\bar r} & 0\cr 
-{\bar t^*_1} & -{\bar t^*_2} & 1 \end{pmatrix}
\begin{pmatrix} \varphi'_1 \cr \varphi'_2 \cr \varphi'_3 \end{pmatrix}
= \begin{pmatrix} 1 & 0 & {\bar t}_1 \cr 0 & 1 & {\bar t}_2 \cr 0 & 0 & 0 \end{pmatrix}
\begin{pmatrix} \varphi_1 \cr \varphi_2 \cr \varphi_3 \end{pmatrix},
\end{eqnarray}
with identification
${\bar s} = r/(sr-q^*q)$,  ${\bar  q} = -q/(sr-q^*q)$, ${\bar  r} = s/(sr-q^*q)$,
${\bar  t}_1 = c$, and ${\bar t}_2 = d$.

This is obviously dual to the case of $m=2$, ${\rm rank}(S)=2$.
Now the presence of $\delta'$-like connection between all lines are guaranteed, 
and the presence or absence of $\delta$-like component is controlled by $c$ and $d$.
The transmission amplitudes, given by
\begin{eqnarray}
\label{e355}
&&\!\!\!\!\!\!\!\!\!\!\!\!
{\cal T}_{31}(k) = \frac{2 {\textrm i} k (c^*s+d^*q^*) - 2 c^*(sr-q^*q)  }
{ k^2+{\textrm i}kF_1 + F_0 }   ,
\nonumber \\
&&\!\!\!\!\!\!\!\!\!\!\!\!
{\cal T}_{12}(k) = \frac{2 {\textrm i} k q + 2 c d^*(sr-q^*q)  }
{ k^2+{\textrm i}kF_1 + F_0 }   ,
\nonumber \\
&&\!\!\!\!\!\!\!\!\!\!\!\!
{\cal T}_{23}(k) = \frac{2 {\textrm i} k (c q^*+d r) - 2 d(sr-q^*q)  }
{ k^2+{\textrm i}kF_1 + F_0 }  ,
\end{eqnarray}
where we set
\begin{eqnarray}
\label{e3551}
&&\!\!\!\!\!\!\!\!\!\!\!\!
F_0= -(sr-q^*q)(1+c^*c+d^*d)   ,
\nonumber \\
&&\!\!\!\!\!\!\!\!\!\!\!\!
F_1 = s+r+c^*c s+ c^* d q + d^* c q^* + d^* d r,
\end{eqnarray}
corroborate this assertion with high energy blockade ${\cal T}_{ij}(\infty)=0$ 
for all $i$ and $j$, and also with the zero energy expressions
${\cal T}_{31}(0)\propto c^* $, ${\cal T}_{12}(0)\propto d^*c$ 
and ${\cal T}_{23}(0)\propto d $.
A numerical example of this case is shown in Fig. 9.
\begin{figure}[h]
  \centering
  \includegraphics[width=9cm]{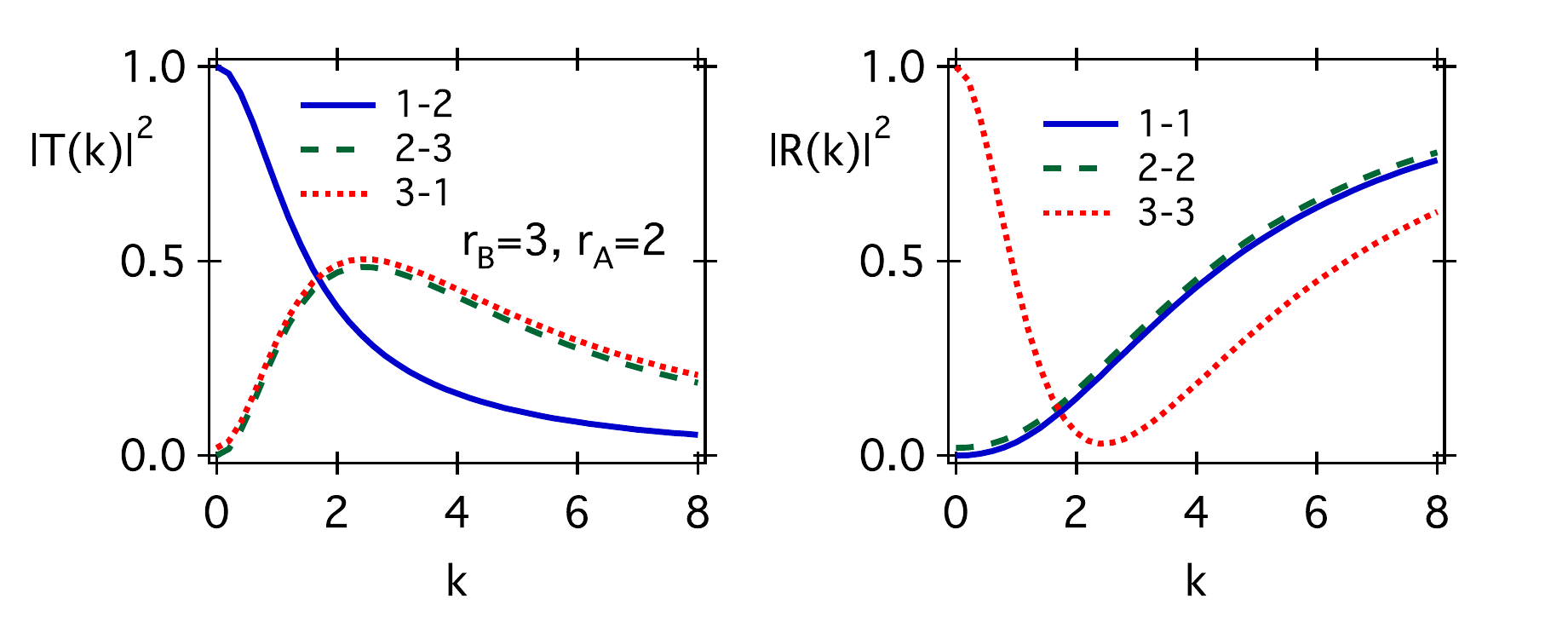}
  \caption{Transmission and reflection probabilities for Y-junction 
  with ${\rm rank}(B)=3$, ${\rm rank}(A)=2$. 
  Parameter values $s_{11}=s_{12}=s_{22}=s_{33}=1$, $s_{13}=s_{23}=2$
  are used in \eqref{e34}.}
  \label{fig:m3s2}
\end{figure}
\subsubsection {\bf rank($B$)=3, rank($A$)=3}
When the ranks of the matrices $A$ and $B$ are both equal
to $n=3$, we have the generic connection 
condition for a quantum particle residing on a joint three lines, namely the
combinations of $\delta$ and $\delta'$ interactions.
\begin{figure}[h]
  \centering
  \includegraphics[width=9cm]{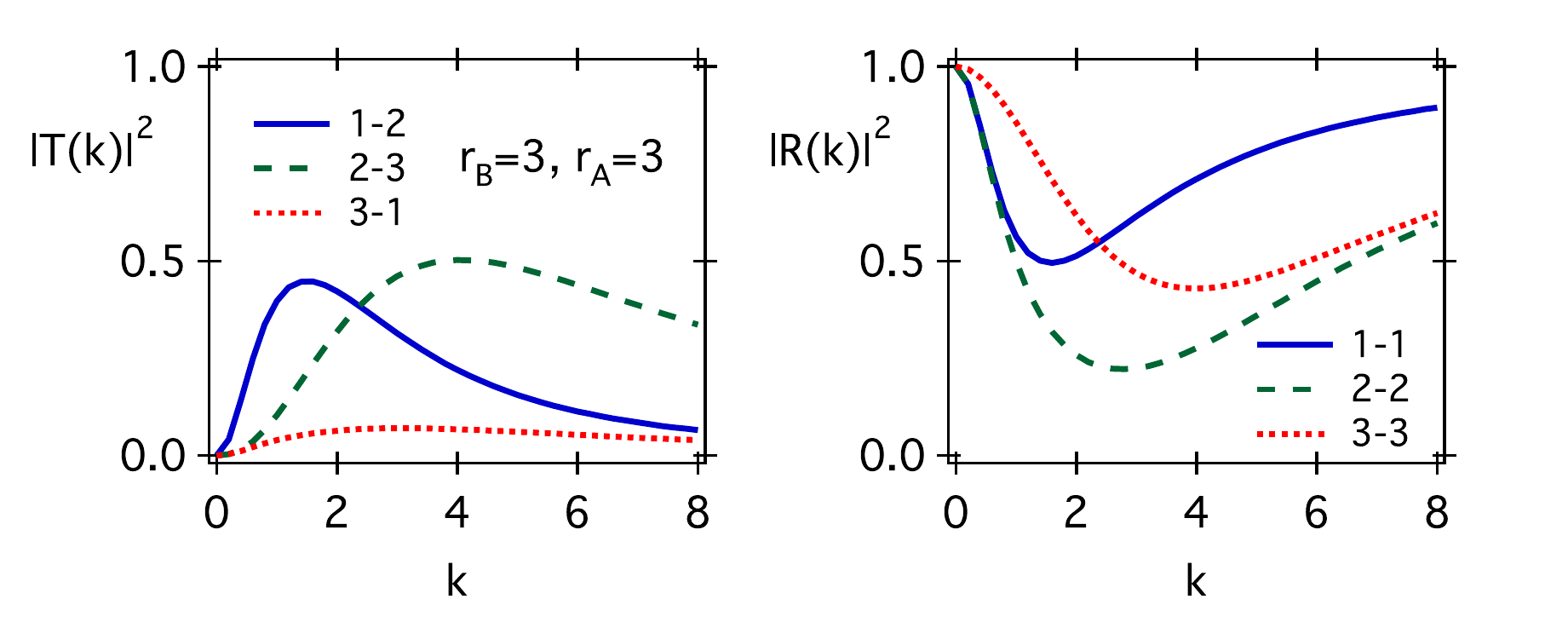}
  \caption{Transmission and reflection probabilities for Y-junction 
  with ${\rm rank}(B)=3$, ${\rm rank}(A)=3$, the generic condition.  
   In the left side, solid line represents $|{\cal T}_{12}(k)|^2$, 
   dashed line $|{\cal T}_{23}(k)|^2$, and dotted line $|{\cal T}_{31}(k)|^2$.
   In the right, solid line represents $|{\cal R}_{1}(k)|^2$, 
   dashed line $|{\cal R}_{2}(k)|^2$, and dotted lin $|{\cal R}_{3}(k)|^2$.
   Parameter values $ s_{11}=-1/3$, $s_{12}=-1$, $s_{13}=1$, 
  $s_{22}=1$, $s_{23}=-3$, $s_{33}=-4$ are used in \eqref{e34}.}
  \label{fig:m3s3}
\end{figure}
Let us look at the transmission amplitudes, which are given by
\begin{eqnarray}
\label{e41}
{\cal T}_{ij}(k) = \frac{-2 {\textrm i} k^2 s_{ij} + 2k \det[{\cal S}_{ji}] }
{k^3 + {\textrm i} k^2 {\rm tr}[S] - k \sum_i\det[{\cal S}_{ii}] -{\textrm i} \det[S] }.
\end{eqnarray}
We have ${\cal T}_{ij}(0) = {\cal T}_{ij}(\infty)=0$ for all $i\ne j$ signifying the
guaranteed presence of both $\delta$-like and $\delta'$-like
components in the connections between all lines.

This expression, along with the analogous expression for $n=r_A=r_B=2$ case,
invites an easy straightforward extension to general $n$. 
A numerical example of this case is shown in Fig. 10.

%

\section {Conclusion}

Our main finding in this article on quantum Y-junction is the fact that
the couplings between each pair of outgoing lines are individually tunable.
The ST form of vertex boundary condition,
which gives the prescription for minimal construction of singular vertex
as a limit of finite potentials, is also found to be instrumental  in identifying
the type of coupling between all pairs of outgoing lines.
Crucial quantity to identify the physics of singular vertex is to be found in
the rank of matrices $A$ and $B$ appearing in the ST form.

Specifically, the pure $\delta$-type coupling is constructed 
from ${\rm rank}(B)=1$  boundary condition, while the pure $\delta'$-type 
coupling is constructed from ${\rm rank}(A)=1$.

Boundary condition corresponding to ST form for $n=3$
with ${\rm rank}(A)= {\rm rank}(B)= 2$ is identified as containing  Y-junction 
with both $\delta$--$\delta$--$\delta'$ type and $\delta'$--$\delta'$--$\delta$ type
singular verteces as limiting cases of parameter values ${\bar t_i}=0$ 
and $t_i=0$, respectively.
Spectral filtering of quantum waves is achieved by these types of singular
vertices.

The extension of our treatment to quantum singular vertex of degree $n=4$,
or ''X-junction'', and then to that with higher $n$ appears tedious, but 
is within reach once the need of detail analysis is required as
a model of quantum single electron devices.
We hope that this work becomes a stepping stone for such extensions.
Obviously, the experimental realization and demonstration with quantum
wires and quantum dots are  highly desired.
Designing real-world approximation for singular vertex of quantum graph
then becomes crucial \cite{CET09,CS98,ET07,KU04}.
\\

We acknowledge the financial support by the Ministry of Education, Culture, Sports, Science and Technology, Japan (Grant number 21540402), and also by the Czech Ministry of Education, Youth and Sports (Project LC06002).
%
%


\end{document}